\documentclass[10pt,b5paper,twoside]{padeu}  %%%% do not change this line
\usepackage{epsfig,natbib_padeu}             %%%% do not change this line

\begin{document}

\title{On the luminosity-redshift relation in brane-worlds with cosmological constant}
\titlerunning{Luminosity-redshift relation in brane-worlds}

\author{Botond Nagy\inst{1},
Zolt\'{a}n Keresztes\inst{2}}
\authorrunning{B. Nagy, Z. Keresztes}
\institute{Departments of Theoretical and Experimental Physics, University of Szeged, 6720 Szeged, D\'{o}m t\'{e}r 9, Hungary
}
\email{\inst{1},Nagy.Botond@stud.u-szeged.hu 
       \inst{2}zkeresztes@titan.physx.u-szeged.hu}

%________________________________________________________________

\abstract{
In this paper we calculate the luminosity distance - redshift relation for a special type of
flat Friedmann brane with cosmological constant. This special case is singled out by its simplicity, the luminosity distance being given in terms of elementary functions. We compare our analytical result with the expresssion of the luminosity distance for the flat Friedmann-Lemaitre-Robertson-Walker (FLRW) universe and discuss the differences.
}

\maketitle

%________________________________________________________________

\section{Introduction}

[\citet{RS}] suggested a new model for the gravitational interaction acting in five non-compact dimensions, the fifth dimension being warped. In brane cosmological models, emerging as generalizations of [\citet{RS}], our observable universe is a four-dimensional space-time hypersurface (the brane), which has cosmological symmetries and is embedded in the warped five-dimensional bulk. The standard model interactions are confined to the brane, but gravitational dynamics is modified as compared with general relativity, at least at high energies (also at late times in the so-called induced gravity models). Consequently, the luminosity - redshift relation is also changed. 

The relation between the luminosity distance and redshift is a powerful tool of the cosmology 
and has a long history of its own [\citet{perl,pad}]. In general relativity, a milestone was the work of Mattig [\citet{matt}], in which this relation has been derived for the FLRW universe with vanishing cosmological constant. 

In Section 2 we discuss various luminosity distance - redshift relations. Subsection 2.1 contains the definition of the radial coordinate distance. In Subsection 2.2 we use a standard method for the calculation of the luminosity 
distance [\citet{star}] in a FLRW universe with cosmological constant. 
The result cannot be represented by elementary functions as it contains elliptic integrals of the first kind.

In Subsection 2.3 we calculate the luminosity distance - redshift relation for the flat Friedmann brane embeded in $Z_{2}$ symmetrically into the five-dimensional Schwarzschild-anti de Sitter space-time (5D SADS). For a special value of the brane tension, this relation becomes even simpler than in general relativity, containing only elementary functions. We briefly discuss the assumptions which led to this special case. We compare  
the luminosity distance - redshift relations for flat Friedmann brane and for FRLW universe with cosmological constant in the Concluding Remarks.
   
\section{Luminosity distance - redshift relations}

We define the luminosity distance [\citet{pad}] in terms of the luminosity $\mathcal{L}$ and the flux $\mathcal{F}$ as:
\begin{equation}
d_{L} (z)=\Big(\frac{\mathcal{L}}{4\pi \mathcal{F}}\Big)^{\frac{1}{2}}=a_{0}(\eta_{0}-\eta) (1+z) \ .
\end{equation}
Here $a_{0}$ represents the value of the scale factor at present time, $\eta_{0}-\eta$ is the radial coordinate distance of the source, and $z$ is the redshift. In order to find the luminosity distance - redshift relation, first we need to calculate the radial coordinate distance.

\subsection{The radial coordinate distance}

Current observational data indicates [\citet{lid}] that the universe is spatially flat. Thus, in this subsection we calculate the radial coordinate distance for the spatially flat Friedmann metric:
\begin{equation}
ds^{2}=-c^{2}d\tau^{2}-a^{2}(\tau)[d\eta^{2}-\eta^{2}(d\theta^{2}+sin^{2}\theta d\varphi^{2})] \ .
\end{equation}
Light rays, perceived by Earth-based observers, travel along null radial geodesics:
\begin{equation}
ds^{2}=d\theta=d\varphi=0 \ .
\end{equation}
Using (3) we can express the radial distance from the metric:
\begin{equation}
\eta_{0}-\eta=\int_{\eta}^{\eta_{0}}d\eta=\int_{t}^{t_{0}} \frac{c\, d\tau}{a(\tau)}=\int_{a}^{a_{0}}\frac{c\, da}{a^{2}H(a)}
\ .\end{equation}
In the last equality we changed from the time variable $t$ to the scale factor $a$ as integration variable and $H$ 
denotes the Hubble parameter. In the above formula the evolution of Hubble parameter is different in the FLRW universe and for a Friedmann brane.

\subsection{The luminosity distance-redshift relation for flat FLRW universe with cosmological constant}

For the flat FLRW, the Friedmann equation which gives to the evolution of the Hubble parameter is

\begin{equation}
H^{2}=\Big(\frac{\dot a}{a}\Big)^{2}=\frac{\kappa^2\rho}{3}+\frac{\Lambda}{3} \label{Fried1}\ ,
\end{equation}
where $\rho$ denotes the density of matter, $\Lambda$ the cosmological constant, and $\kappa^{2}=8\pi G$. If we divide this equation with the square of the Hubble constant (the present value of the Hubble parameter) $H_{0}^{2}$, we obtain:
\begin{equation}
\frac{H^2}{H_0^2}=\Omega_{\rho}\frac{a_0^3}{a^3}+\Omega_{\Lambda}\ ,
\end{equation}
where we have introduced
\begin{equation}
\Omega_{\rho}=\frac{\kappa^2\rho_{0}}{3H_{0}^{2}}\ ,
\end{equation}
\begin{equation}
\Omega_{\Lambda}=\frac{\Lambda}{3H_{0}^{2}}\ .
\end{equation}
Knowing the evolution of the Hubble parameter, Eq. (\ref{Fried1}) the luminosity is found as
\begin{equation}
d_L(z)=\frac{c(1+z)}{3^{\frac{1}{4}}H_{0}\Omega_{\rho}^{\frac{1}{3}}\Omega_{\Lambda}^{\frac{1}{6}}} [\textbf{F}(\varphi_{0},\varepsilon)-\textbf{F}(\varphi,\varepsilon)]\ ,
\end{equation}
where
\begin{equation}
\varphi_0=arccos\frac{(1-\sqrt{3}) \Omega_{\Lambda}^{\frac{1}{3}}+\Omega_{\rho}^{\frac{1}{3}}}
                     {(1+\sqrt{3}) \Omega_{\Lambda}^{\frac{1}{3}}+\Omega_{\rho}^{\frac{1}{3}}}\ ,
\end{equation}
\begin{equation}
\varphi=arccos\frac{(1-\sqrt{3}) \Omega_{\Lambda}^{\frac{1}{3}}+\Omega_{\rho}^{\frac{1}{3}}(1+z)}
                   {(1+\sqrt{3}) \Omega_{\Lambda}^{\frac{1}{3}}+\Omega_{\rho}^{\frac{1}{3}}(1+z)}\ ,
\end{equation}
and
\begin{equation} 
\varepsilon=\frac{1}{2}+\frac{\sqrt{3}}{4}\ . 
\end{equation} 
The function $\textbf{F}(\varphi,\varepsilon)$ is the elliptic integral of first kind, with the variable $\varphi$ and the argument $\varepsilon$. We can see that even in the general relativistic case, the luminosity distance - redshift relation can be given only in terms of elliptic functions.

\subsection{Flat Friedmann brane}
The metric on the Friedmann brane is the same as in the case of FLRW universe, thus we can use the previous method for the calculation of the radial coordinate distance. Only the evolution of the Hubble parameter is different for a Friedmann brane embedded symmetrically into the 5D SADS space-time (for the most generic form of this equation see [\citet {lag}]): 
\begin{equation} 
H^2=\frac{\kappa^2 \rho}{3} \Big(1+\frac{\rho}{2\lambda}\Big)+\frac{\Lambda}{3}+\frac{2 \bar{m}}{a^4}\ . 
\end{equation} 
New source terms arise as compared to (\ref{Fried1}) from the assumptions that our universe is a brane and there are identical black holes with mass $\bar{m}$ in both bulk regions. Here $\lambda$ is the brane tension. We introduce the following notations: 
\begin{equation} 
\Omega_{\lambda}=\frac{\kappa^2\rho_{0}}{6\lambda H_{0}^{2}}\ , 
\end{equation}
\begin{equation} 
\Omega_{d}=\frac{2\bar m}{a_{0}^{4} H_{0}^{2}}\ . 
\end{equation} 
The radial distance, after a short rearrangement, is
\begin{equation} 
\eta_0-\eta=\frac{c}{H_0\Omega_{\Lambda}^{\frac{1}{2}}}\int_{a}^{a_0}\frac{a\, da}{\Big[\Big(a^3+\frac{\Omega_{\rho} a_0^3}{2\Omega_{\Lambda}}\Big)^2 +\Big(\frac{\Omega_{\lambda}}{\Omega_{\Lambda}}-\frac{\Omega_{\rho}^2}{4\Omega_{\Lambda}^2}\Big)a_0^6+\frac{\Omega_{d}}{\Omega_{\Lambda}}a_0^4 a^2\Big]^{\frac{1}{2}}}\ .
\end{equation} 
In general, this integral leads to elliptic functions, however in the special case, when both the second and the third terms in the denominator vanish, is can be given in terms of elementary functions. For this, two conditions have to be satisfied. The first is 
\begin{equation} 
\Omega_{d}=0\ . 
\end{equation} 
This assumption is realistic at late times because $\Omega_{d}$ is proportional to $a_0^4$. Its direct implication is that the brane, rather than being embedded into the 5D SADS space-time, is embedded into a five-dimensional anti de Sitter (5D ADS) bulk. The second assumption is
\begin{equation} 
\Omega_{\lambda}=\Omega_{\rho}^2/(4\Omega_{\Lambda})\ , 
\end{equation} 
or equivalently
\begin{equation} 
\kappa^{2}\lambda=2\Lambda\ . 
\end{equation} 
As observational evidence suggest $\Omega_{\rho}=0.27$ and for flat universe, we have $\Omega_{\Lambda}$+$\Omega_{\lambda}$+$\Omega_{\rho}=1$, a quadratic equation for $\Omega_{\lambda}$ emerges. Both solutions of this quadratic equation are positive. The values of $\Omega_{\Lambda}$ and $\lambda$ are collected in \textit{Table 1}. 
\begin{table} 
   \caption{The values of $\Omega_{\Lambda}$ and brane tension.} 
   \label{tab:example} 
   \begin{center} 
   \begin{tabular}[!h]{cc} 
   \hline 
   \hline 
   \noalign{\smallskip} 
    $ \Omega_{\Lambda}$ & $\lambda (10^{-60}TeV^{4})$ \\ 
   \noalign{\smallskip} 
   \hline 
   \noalign{\smallskip} 
       0.704 & 38.375     \\  
       0.026 & 1.4173 \\  
   \noalign{\smallskip} 
   \hline 
   \noalign{\smallskip} 
   \end{tabular} 
   \end{center} 
   \end{table}
\begin{figure}[!h] 
   \centering 
   \includegraphics[width=0.40\linewidth]{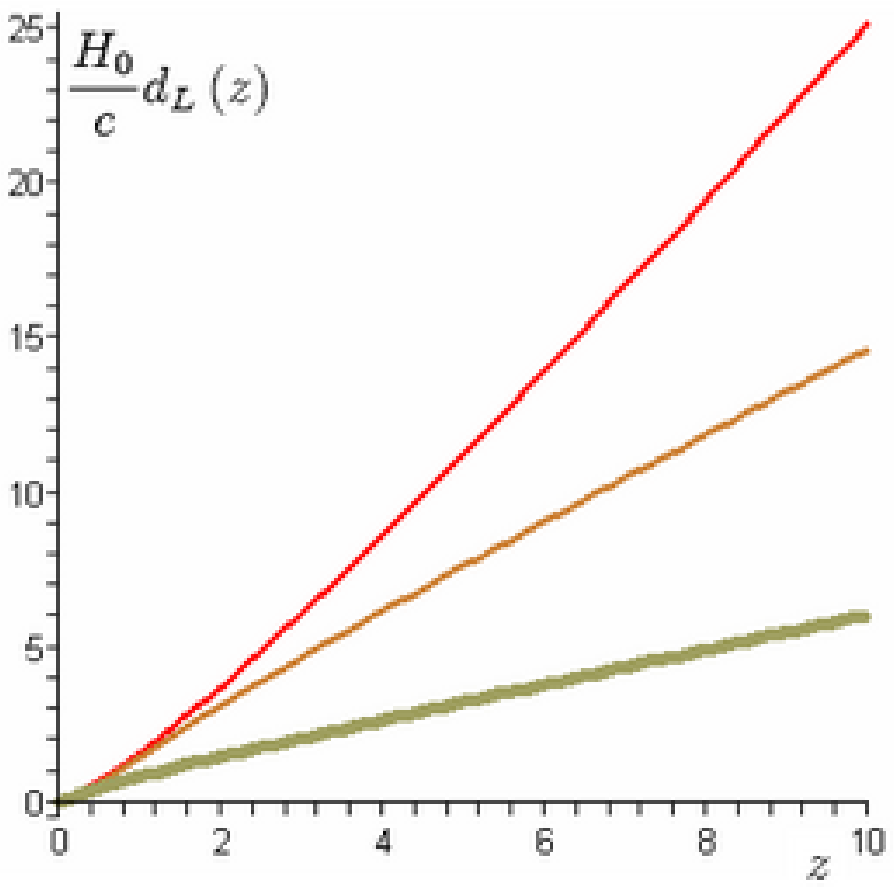}
    \hskip1.5cm 
   \includegraphics[width=0.40\linewidth]{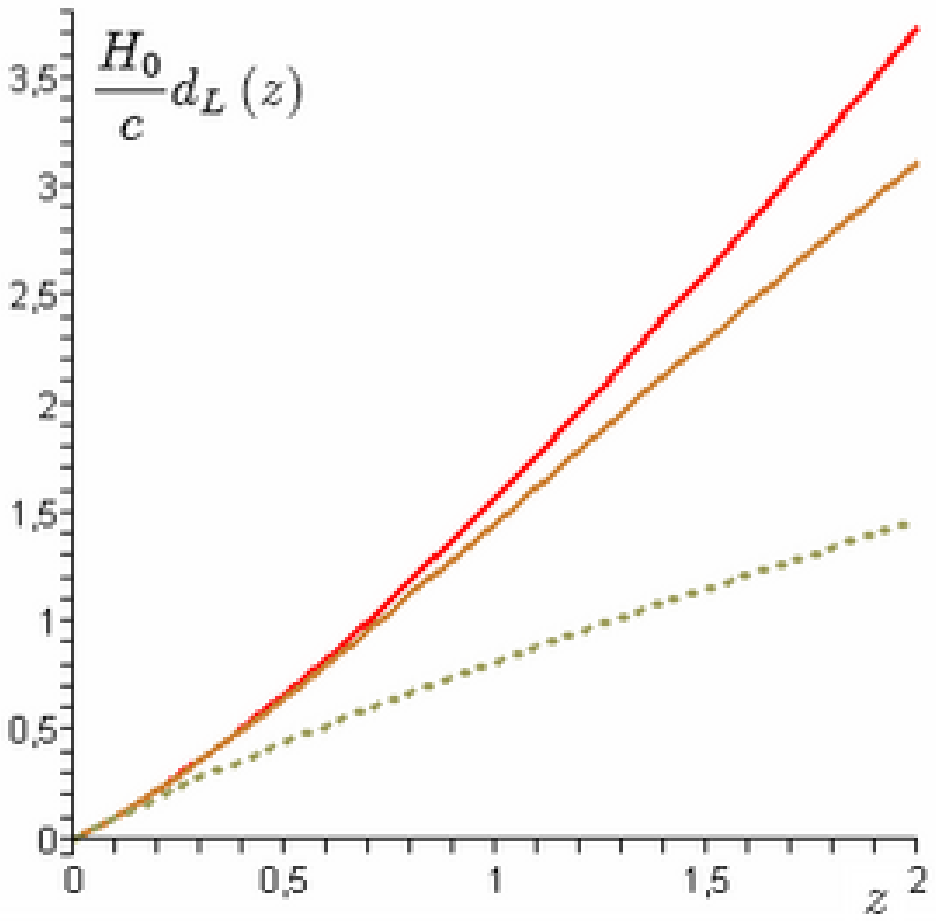} 
   \caption{{\it Left:} The luminosity distance plotted as function of redshift in the range $z=0..10$. 
The upper curve represents the solution for FLRW universe, the middle and the lowest are for the Friedmann brane 
with $\Omega_{\Lambda}$=$0.704$ and $0.026$, respectively.     
            {\it Right:} The same, in the range $z=0..2$.} 
   \label{Fig1} 
   \end{figure} 
We note that these values of the brane tension are much below the minimal value of $\lambda$ predicted to be $1(TeV)^{4}$ [\citet{maar}]. 
However with the conditions (17) and (18) satisfied, the luminosity distance has a very simple expression:
\begin{eqnarray}
\lefteqn{d_L=\frac{1}{6} \frac{c(1+z)}{2^{\frac{-1}{3}}H_{0}\Omega_{\rho}^{\frac{1}{3}}\Omega_{\Lambda}^{\frac{1}{6}}} 
\Big\{ln\frac{(1-h+h^2)[1+h(1+z)]^2}{[1-h(1+z)-h^2(1+z)^2](1+h)^2}+{}}\nonumber\\  &&{}
+2\sqrt{3}Big[arctan\frac{\sqrt{3}}{3}\Big(\frac{2}{h}-1\Big)
             -arctan\frac{\sqrt{3}}{3}\Big(\frac{2}{h}(1+z)-1\Big)Big]\Big\}\ \ ,
\end{eqnarray}
where we have introduced: 
\begin{equation} 
h=\Big(\frac{\Omega_{\rho}}{2\Omega_{\Lambda}}\Big)^{\frac{1}{3}}\ . 
\end{equation}

\section{Concluding remarks} 
We have derived the analytical expressions of the luminosity distances for both a flat FLRW universe with cosmological constant and a Friedmann brane embedded into 5D ADS bulk. These expressions are substantially different, as they depend on the Friedmann equation. In the case of the Friedmann brane we have imposed two simplifying assumptions yielding the luminosity distance in terms of elementary functions. There are two values of the cosmological constants and of the brane tension, which are in accordance with these assumptions. The higher value of the $\Omega_{\Lambda}$ (see: \textit{Table 1}) is very close to today's preferred value [\citet{lid}]. 

The luminosity distances as function of redshift for all three cases is represented in Fig. 1. On the two plots, $d_{L}$ is represented from $z=0$ to $z=10$ and $z=0$ to $z=2$, respectively. The motivation for the second graph is that supernova observations extend nowadays up to $z=2$. On the plots, we see that all three luminosity distances grow monotonically with increasing redshift. The steepest curve belongs to the FLRW universe. The middle curve is for the case of the brane with the higher value of the cosmological constant. This curve, in the range $z=0..2$, is extremely close to one pertinent to a flat FLRW universe. 

Since the values of both brane tensions are much below the theoretically predicted limit, our brane model qualifies as a "toy model". The constraints on brane tension [see: \citet{maar}] imply that $\Omega_{\lambda}$ should be small. Nowadays, $\Omega_{d}$ is also small, being proportional to $a_0^4$. Thus a perturbative treatment can give rise to a more realistic solution for the luminosity distance for Friedmann brane models [see: \citet{KN}]. However, such realistic solutions for the luminosity distance will be more complicated than the correponding expressions in general relativity.

%________________________________________________________________
\begin{acknowledgement}
We thank L\'{a}szl\'{o} \'{A}rp\'{a}d Gergely for raising this problem and
guidance in its elaboration. This work was supported by OTKA grants no.
T046939 and TS044665.
\end{acknowledgement}
%________________________________________________________________

\end{document}